\documentclass[prb,twocolumn,showpacs]{revtex4}

\usepackage{graphicx}
\usepackage{amsmath}
\usepackage{amsfonts}
\usepackage{bm}
\usepackage{bbm}

\begin{document}


\title{Polarization-anisotropy induced spatial anisotropy of
polariton amplification in planar semiconductor microcavities}

\author{Stefan Schumacher}

\affiliation{College of Optical Sciences,
             University of Arizona,
             Tucson, Arizona 85721, USA}

\date{\today}

\newcommand{\ME}{{\genfrac{}{}{0.0pt}{3}{\text{TM}}{\text{TE}}}}
\newcommand{\EM}{{\genfrac{}{}{0.0pt}{3}{\text{TE}}{\text{TM}}}}
\newcommand{\TM}{\text{TM}}
\newcommand{\TE}{\text{TE}}
\newcommand{\probe}{\text{probe}}
\newcommand{\pump}{\text{pump}}
\newcommand{\probepump}{{\genfrac{}{}{0.0pt}{3}{\text{probe}}{\text{pump}}}}

\newcommand{\XY}{{\genfrac{}{}{0.0pt}{3}{\text{X}}{\text{Y}}}}
\newcommand{\YX}{{\genfrac{}{}{0.0pt}{3}{\text{Y}}{\text{X}}}}
\newcommand{\X}{\text{X}}
\newcommand{\Y}{\text{Y}}


\begin{abstract}
Based on a microscopic many-particle theory we investigate the
amplification of polaritons in planar semiconductor microcavities.
We study a spatially perfectly isotropic microcavity system and
excitation geometry. For this system, our analysis shows that a
pump-induced vectorial polarization anisotropy can lead to a spatial
anisotropy in the stimulated amplification of polaritons. This
effect is brought about by the interplay of the
longitudinal-transverse cavity mode splitting (TE-TM splitting) and
the spin-dependence of the polariton-polariton scattering processes.
\end{abstract}

\maketitle

\textsc{Introduction} -- In the past decade the parametric
amplification of polaritons in planar semiconductor microcavities
has been intensively investigated in theory and experiment, see,
e.g.,
Refs.~\onlinecite{Savvidis2000,Huang2000,Ciuti2000,Stevenson2000,Saba2001,Whittaker2001}
or the reviews given in
Refs.~\onlinecite{Ciuti2003,Baumberg2005,Keeling2007}. Whereas the
polariton amplification was initially reported in a co-circularly
polarized pump-probe setup\cite{Savvidis2000}, for other
configurations it can show a rich vectorial polarization
dependence.\cite{Lagoudakis2002,Kavokin2003,Eastham2003,Renucci2005,Kavokin2005,Dasbach2005,Krizhanovskii2006,Sanvitto2006,Klopotowski2006,Schumacher2007a,Martin2007}
This vectorial polarization dependence is governed by the underlying
spin-dependent polariton-polariton scattering processes,
longitudinal-transverse cavity mode splitting (TE-TM splitting), and
possible structural anisotropy of the investigated systems.

In the most intensively studied configuration, a pump pulse induces
a coherent polariton density close to the inflection point of the
lower polariton branch (LPB), the so-called ``magic angle''. In this
configuration phase-matched resonant pairwise scattering of
pump-induced polaritons into probe and four-wave mixing (FWM)
directions is allowed which makes the polariton amplification most
efficient. However, for oblique pump excitation the intrinsic
cylindrical symmetry of the planar microcavity system is destroyed
by the optical fields.

\begin{figure}[b]
\includegraphics[scale=1.0]{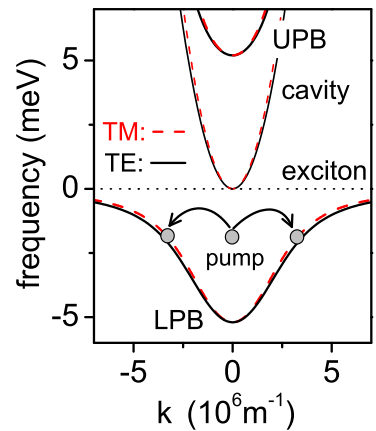}
\caption{\label{dispersion}(color online) Shown are the isotropic TE
(solid) and TM (dashed) cavity polariton modes on the lower (LPB)
and upper (UPB) polariton branch, respectively. Also shown are the
bare cavity and exciton dispersions. The basic pairwise off-axis
scattering of pump-induced polaritons is indicated. For the GaAs
parameters used and for the modeling of the cavity dispersion see
Ref.~\onlinecite{Schumacher2007a}.}
\end{figure}

In this paper, we study an excitation geometry where the incoming
pump does not necessarily spoil the symmetry of the system. As the
simplest such configuration, we investigate the polariton
amplification for excitation with a pump pulse, which is incident
along the system's symmetry axis, i.e., perpendicular to the plane
of the embedded quantum well along the $z$ axis. In this excitation
configuration the pump-excited polaritons carry zero in-plane
momentum $\mathbf{k}_p=0$. As illustrated in Fig.~\ref{dispersion},
for pump frequency above the LPB, FWM processes triggered by
fluctuations in the cavity photon field can give rise to pairwise
off-axis scattering of pump-excited polaritons into two polaritons
with finite and opposing in-plane momentum $\mathbf{k}$ and
$-\mathbf{k}$. For a pump-induced exciton density above the
stimulated amplification threshold these scattering processes can
lead to strong off-axis ($\mathbf{k}\neq0$) signals as has been
demonstrated in Ref.~\onlinecite{Romanelli2007} for a slightly
different excitation geometry. For the perfectly isotropic system
studied in this work, spontaneous (fluctuation-induced) off-axis
pattern formation is expected which is invariant under rotation
about the propagation direction of the pump.

Based on a microscopic many-particle theory we discuss that even if
the incoming optical pump field does not destroy the spatial
isotropy of the system, a specific choice of its vectorial
polarization state can lead to a pronounced azimuthal angular
dependence of the spontaneous signals in the off-axis directions. We
discuss the interplay of the longitudinal-transverse cavity mode
splitting (TE-TM splitting) and the spin-dependent
polariton-polariton scattering processes that leads to this spatial
anisotropy, solely induced by a vectorial polarization anisotropy.

\textsc{Microscopic theory} -- Our theoretical analysis is based on
the microscopic many-particle theory presented in
Ref.~\onlinecite{Schumacher2007a}. It includes spin and frequency
dependent polariton-polariton
scattering\cite{Kwong2001a,Kwong2001b,Savasta2003,Schumacher2006}
and a TE-TM cavity mode splitting in the coupled cavity-confined
field and exciton dynamics. Calculations are done for the same GaAs
based microcavity system studied in
Ref.~\onlinecite{Schumacher2007a} but for a slightly smaller
intrinsic dephasing for the excitonic polarization amplitude and the
fields in the cavity modes, $\gamma_x=\gamma_c=0.15\,\mathrm{meV}$.
Within this framework and for steady-state monochromatic pump
excitation with frequency $\omega_p$, the linearized coupled
dynamics for field and excitonic polarization components with
in-plane momenta $\mathbf{k}$ and $-\mathbf{k}$ can be cast into the
following simple form:
\begin{align}\label{eqlinstab}
\hbar\dot{\tilde{\mathbf{p}}}_{\mathbf{k}}=M_{\mathbf{k}}\tilde{\mathbf{p}}_{\mathbf{k}}\,.
\end{align}
The vector $\tilde{\mathbf{p}}_{\mathbf{k}}$ groups together the
dynamic variables for field components (in the TE and TM cavity
modes, respectively) and excitonic polarization components
(longitudinal or transversal to $\mathbf{k}$, respectively).
Analyzing the linear stability of the pump-driven polariton dynamics
against off-axis fluctuations, no incoming fields in the
$\mathbf{k}$ and $-\mathbf{k}$ directions are explicitly considered.
The matrix $M_{\mathbf{k}}$ is a $\mathbf{k}$-dependent but time
independent matrix where all the system and pump parameters enter.
The components of $\tilde{\mathbf{p}}_{\mathbf{k}}$ carry only the
time dependence relative to the time dependence
$\sim\mathrm{e}^{-i\omega_p t}$ that is imposed by the pump field.
This additional time dependence is determined by the eigenvalues
$\lambda$ of $M_{\mathbf{k}}$. The two-exciton continuum is treated
in Markov approximation as in Ref.~\onlinecite{Schumacher2007a} but
non-Markovian contributions from excitation of the bound biexciton
state are included in Eq.~(\ref{eqlinstab}). This Markov
approximation should be particularly well fulfilled for the present
study since the dominant and thus relevant modes oscillate with or
close to the pump frequency $\omega_p$ (cf. Fig.~\ref{dispersion}).
We neglect the influence of excitonic phase-space filling
nonlinearities which have been found to be insignificant for the
stimulated polariton amplification.\cite{Schumacher2007a} To
concentrate on the issues under investigation, we do not consider
possible complications of our discussion from
bistability\cite{Wouters2007} or multistability\cite{Gippius2007}
for the steady-state pump-induced polarization.

\begin{figure}[t]
\includegraphics[scale=0.59]{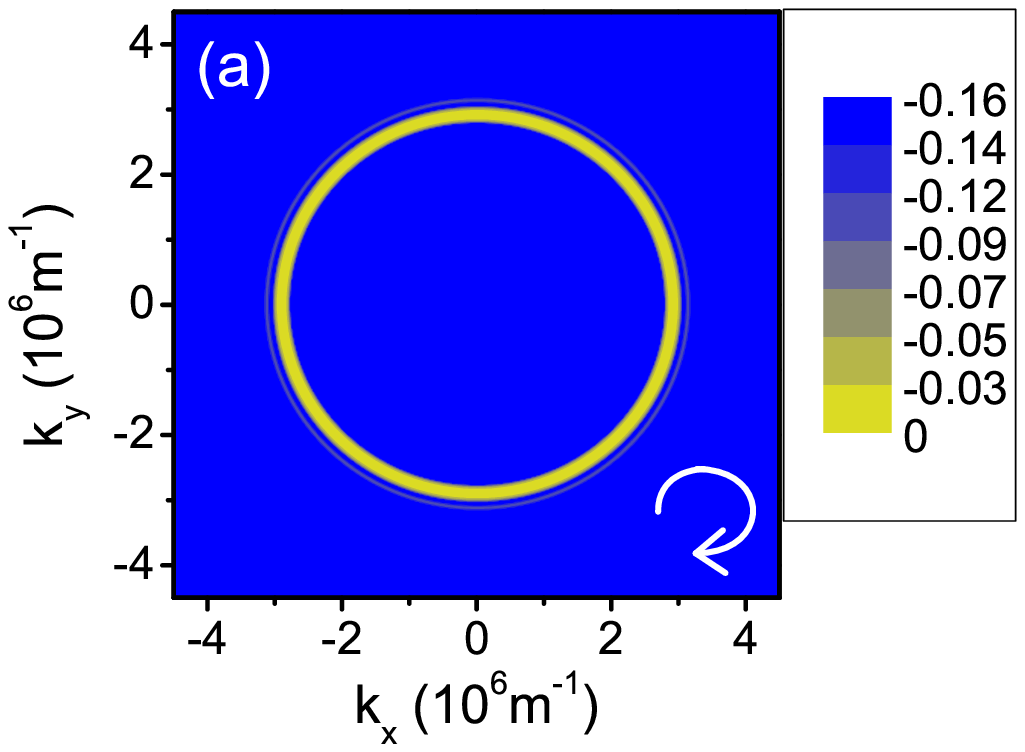}\\[4mm]  \includegraphics[scale=0.59]{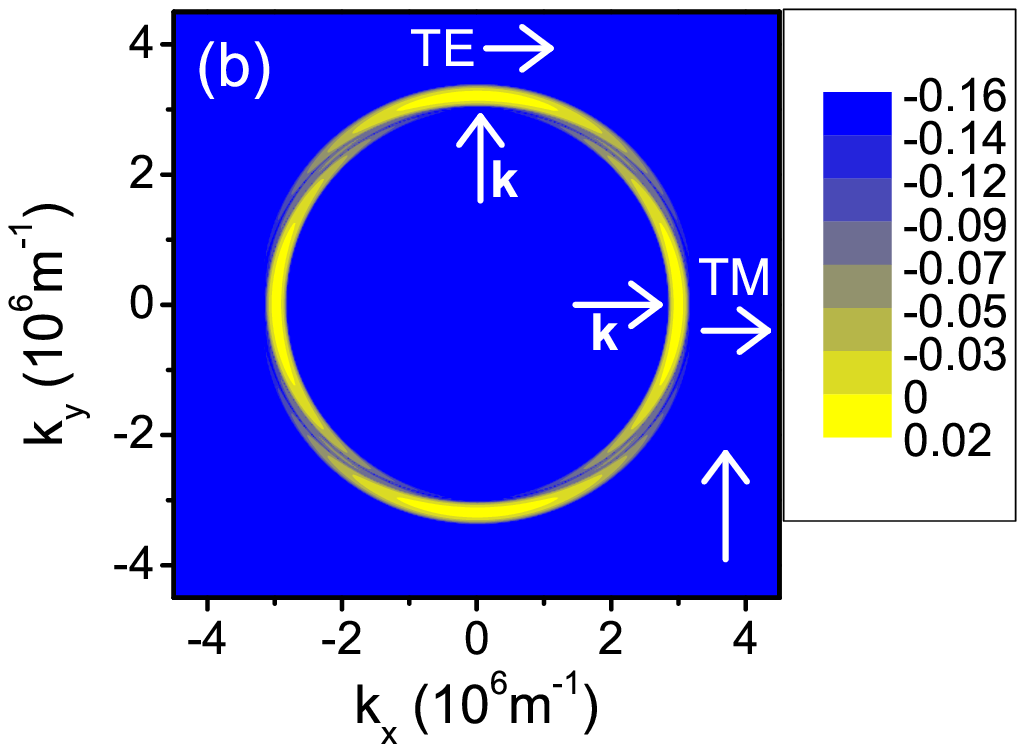}
\caption{\label{anglecirc}(color online) Maximum growth rate for
signals with in-plane momenta $\mathbf{k}$ and $-\mathbf{k}$.
Results are given in $\mathrm{meV}$ and the vectorial pump
polarization state is indicated in the lower right corner of each
panel. Results are calculated including TE-TM cavity-mode splitting
and excitonic correlations. (a) Circularly polarized pump. The
off-axis scattering is isotropic. (b) Linearly polarized pump. The
preferred vectorial polarization state for the signals scattered
off-axis ($\mathbf{k}\neq0$) is perpendicular to the pump's
polarization vector. For $\mathbf{k}$ parallel (perpendicular) to
the pump's polarization vector, scattering in the TE (TM) mode is
preferred. The momentum $\mathbf{k}$ and the preferred polarization
vector for the scattered signals is indicated for $\mathbf{k}$
parallel and perpendicular to the pump's polarization vector,
respectively. Together with the TE-TM mode splitting, this
polarization dependence leads to the spatial anisotropy in the
off-axis signals.}
\end{figure}

Within this framework, the eigenvalues $\lambda$ of $M_{\mathbf{k}}$
contain comprehensive information about temporal growth or decay
(real part of $\lambda$) and frequency (imaginary part of $\lambda$)
of the possible off-axis modes as long as the system dynamics is
linear in the off-axis contributions. Information about the
vectorial polarization state of these modes is contained in the
respective eigenvectors. We investigate the system dynamics for a
total (including both vectorial polarization states) pump-induced
exciton density of $5\cdot10^9\,\mathrm{cm^{-2}}$, which is close to
the amplification threshold, and pump frequency $\omega_p$ tuned
$2\,\mathrm{meV}$ below the bare exciton resonance. The data
presented in the following quantitatively depend on the choice of
excitation, cavity and other material parameters. However, the
qualitative nature of the results is robust. We diagonalize
$M_{\mathbf{k}}$ for different in-plane momenta $\mathbf{k}$, and
analyze the real-part of the eigenvalue $\lambda$ with the largest
real part. This gives us the growth rate of the initially fastest
growing and therefore ``most unstable'' mode, which will eventually
dominate the system dynamics for sufficiently long growth periods.

\textsc{Results and Discussion} -- For excitation with a circularly
polarized pump, Fig.~\ref{anglecirc}(a) shows the real-part of the
eigenvalue $\lambda$ of $M_{\mathbf{k}}$ with the largest real part,
i.e., the maximum growth (for $\mathrm{Re}\{\lambda\}>0$) or minimum
decay (for $\mathrm{Re}\{\lambda\}<0$) rate for each $\mathbf{k}$.
(In the following loosely referred to as `growth'.) The largest
growth rate is found on the `elastic circle' where pairs of the
off-resonantly excited pump polaritons can resonantly be scattered
on the pump-renormalized polariton dispersion (cf.
Fig.~\ref{dispersion}). No angular dependence of the displayed
growth rates, and thus no anisotropy of the stimulated scattering is
found. Note that for the chosen pump density the stimulated
amplification threshold is not reached (all
$\mathrm{Re}\{\lambda\}<0$) in Fig.~\ref{anglecirc}(a), all modes
are exponentially decaying and no significant off-axis scattering of
light initiated by fluctuations is expected.

\begin{figure}[t]
\includegraphics[scale=0.59]{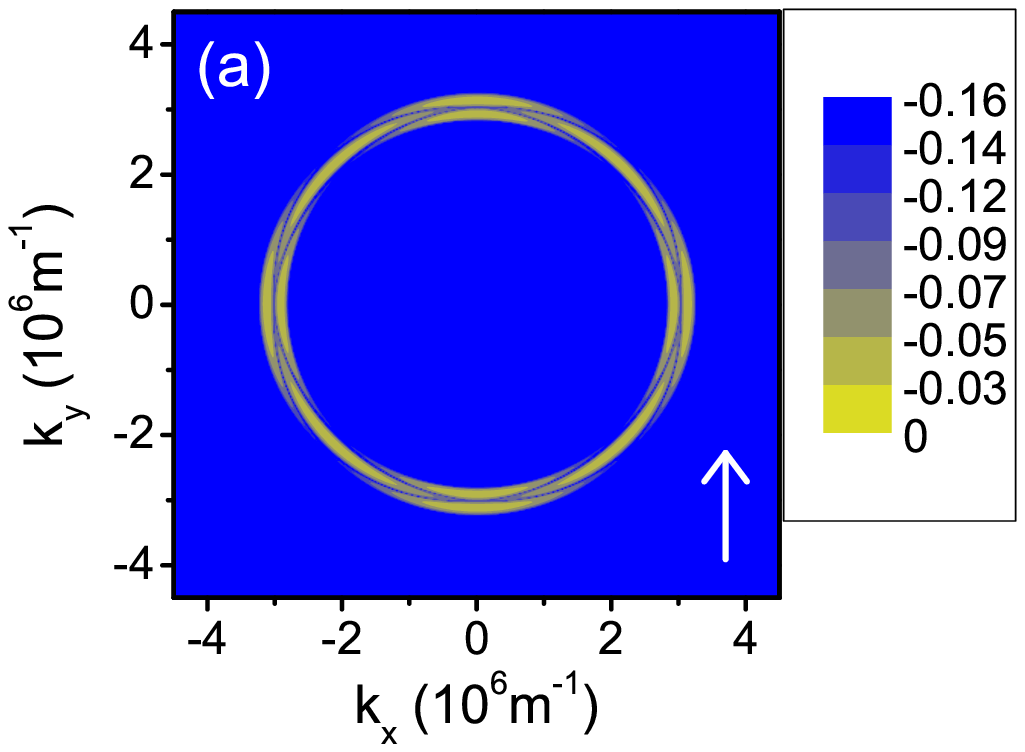}\\[4mm] \includegraphics[scale=0.59]{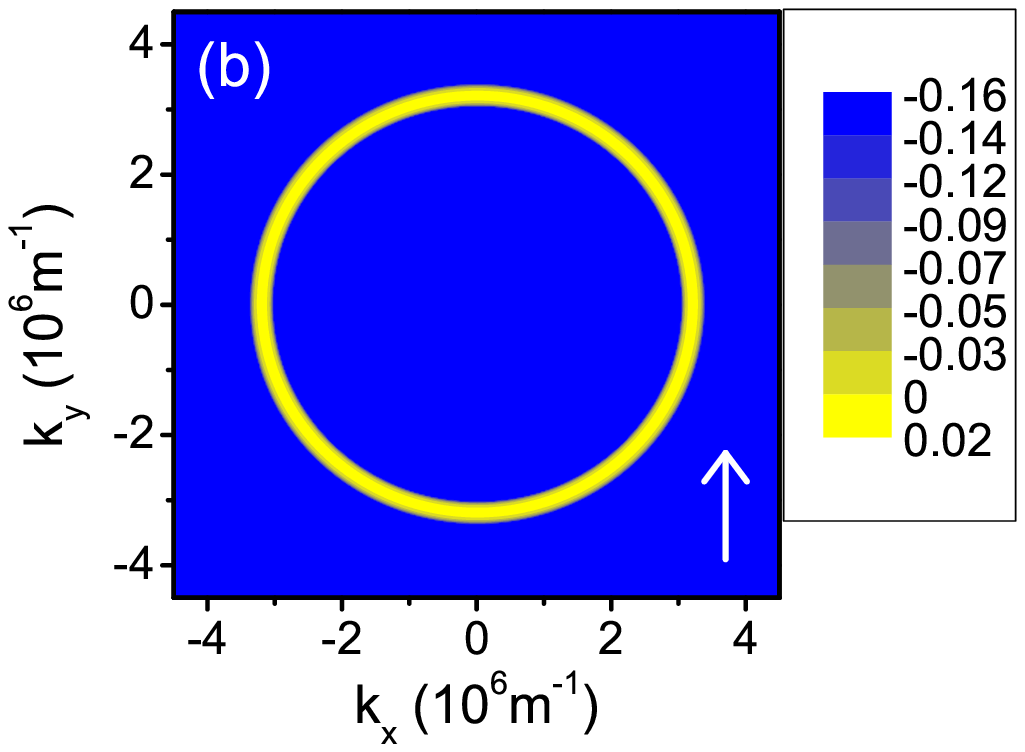}
\caption{\label{angleHF}(color online) (a) Same as
Fig.~\ref{anglecirc}(b) but evaluated on the mean-field Hartree-Fock
level neglecting excitonic correlations. (b) Same as
Fig.~\ref{anglecirc}(b) but without TE-TM cavity-mode splitting.
(a), (b) The vectorial pump polarization state is indicated in the
lower right corner of each panel.}
\end{figure}

This situation changes for excitation with a linearly polarized pump
as shown in Fig.~\ref{anglecirc}(b). By choosing a particular
orientation of the vectorial polarization of the pump, the
spin-dependent polariton-polariton scattering together with the
TE-TM cavity mode splitting induces a spatial anisotropy for the
growth of off-axis signals over time. It is worthwhile to emphasize
again that this anisotropy comes about without any anisotropy in the
cavity or the quantum well. The stimulated scattering intrinsically
depends on the orientation of $\mathbf{k}$ to the pump's linear
polarization vector: Owing to the spin-dependence of the underlying
polariton-polariton scattering processes, polaritons scattered
off-axis are preferably polarized perpendicular to the pump's
vectorial polarization state.\cite{Schumacher2007a} This preference
leads to different scattering rates into the TE or TM polarized
modes, depending on the orientation of the off-axis momenta
$\mathbf{k}$ and $-\mathbf{k}$ of the scattered polaritons relative
to the pump's polarization vector. For the TE mode, the electric
field vector is perpendicular to the plane of incidence, thus also
perpendicular to $\mathbf{k}$. For the TM mode, the component of the
electric field vector in the quantum well plane is parallel to
$\mathbf{k}$. Therefore, for the results shown in
Fig.~\ref{anglecirc}(b), for $\mathbf{k}$ parallel to the pump's
polarization vector, scattering in the TE mode is preferred. For
$\mathbf{k}$ perpendicular to the pump's polarization vector,
scattering into the TM mode is preferred. A similar anisotropy has
been observed earlier but in the linear optical regime, in the
ballistic propagation of cavity polariton wave
packets\cite{Langbein2007} and in Ref.~\onlinecite{Kavokin2005b}
where an optical spin Hall effect was proposed. In
Refs.~\onlinecite{Kavokin2005b,Langbein2007} the TE-TM cavity mode
splitting has led to a four-fold rotation symmetry in the results.
In our case the spin-dependent many-particle interactions reduce the
rotation invariance further to a two-fold rotation symmetry. As
shown in Fig.~\ref{angleHF}(a), a four-fold rotation symmetry in the
polariton amplification can only be regained if the
polariton-polariton scattering is treated on the mean-field
Hartree-Fock level, neglecting excitonic correlations in the
scattering processes. For scattering with $\mathbf{k}$ either
parallel or perpendicular to the pump's polarization vector,
$M_{\mathbf{k}}$ in Eq.~(\ref{eqlinstab}) is block-diagonal.
Consequently, in these directions the polaritons are scattered
off-axis either into the TE or into the TM mode. For all other
orientations, the eigenvectors of $M_{\mathbf{k}}$ contain
contributions from both TE and TM modes. Note that without
correlations in Fig.~\ref{angleHF}(a) again the amplification
threshold is not reached for the chosen density.

Finally, Fig.~\ref{angleHF}(b) shows the growth rate including
correlations but for zero TE-TM cavity mode splitting. Circular
symmetry is recovered and the amplification threshold is reached.
This indicates that minimization or proper engineering of the TE-TM
cavity-mode splitting -- at least close to $\omega_p$ -- can be used
to rebuild the isotropy of the stimulated amplification mechanism.
This way possible benefits of excitation with a linearly polarized
pump, such as a lower threshold density or polarization-selective
separation of the off-axis signals from the pump, respectively, may
be accessible without inducing an undesired anisotropy in the system
dynamics.

Strictly speaking, the linear stability analysis discussed above
only reveals information about the initial growth of off-axis
fluctuations as long as the system dynamics is linear in
the resulting off-axis signals. 
To avoid a clear deviation from this situation in an experimental
attempt to verify the discussed results, the pump frequency
$\omega_p$ should be chosen such that the off-axis scattering of
pump polaritons does not lead to a significant buildup of polariton
density at the ``magic angle''. This precaution is in analogy to
Ref.~\onlinecite{Romanelli2007} and prevents that polariton
scattering at the ``magic angle'' would eventually take over the
nonlinear system dynamics with increasing intensity of the off-axis
signals over time.

As an alternative to the steady-state experiment, all assumptions
entering the linear stability analysis can be well satisfied in a
typical pump-probe experiment: The pump pulse needs to be long
enough, so that the steady-state limit is imitated in a good
approximation. The incoming probe pulse must be weak so that even
after amplification it does not significantly influence the
polarization dynamics in the pump direction. The incoming probe
pulse should also be short enough so that it does not considerably
drive the polarization dynamics, but merely acts as a seed for the
off-axis signal in the probe direction. With the pump pulse in
normal incidence, the probe pulse is used to map out the
amplification in the $k_x,k_y$ plane.

\textsc{Remarks \& Conclusions} -- We have shown that in a perfectly
isotropic microcavity system a vectorial polarization anisotropy of
the pump can induce a spatial anisotropy for the stimulated
amplification of polaritons. This anisotropy is brought about by the
combined influence of a TE-TM cavity mode splitting and the
spin-dependent polariton-polariton scattering processes. Besides its
general interest, this effect may have an influence on, e.g., the
generation of correlated photon pairs in the stimulated
amplification regime.\cite{Romanelli2007} Angular dependencies as
discussed in this work are also expected to play a role in other
systems\cite{Diederichs2006,Schumacher2007d} as long as they are not
overshadowed by structural imperfections.

\textsc{Acknowledgments} -- The author gratefully acknowledges
various fruitful discussions with Nai H. Kwong and Rolf Binder
throughout this project. He also acknowledges support by the
Deutsche Forschungsgemeinschaft (DFG, project No. SCHU~1980/3-1).
This work has further been supported by ONR, DARPA, and JSOP.

\bibliography{../../../literature}

\end{document}